\documentclass[pra,showpacs,twocolumn,aps,superscriptaddress,a4paper]{revtex4-1}
\usepackage{dcolumn,amssymb,amsmath,amsfonts,graphicx,latexsym,color}
\usepackage{epstopdf}
\begin{document}

\title{Three-component topological superfluid in one-dimensional Fermi gases with spin-orbit coupling}

\author{Jie Chen}
\affiliation{Department of Physics, Zhejiang Normal University, Jinhua 321004, China}
 \author{Hui Hu}
\affiliation{Centre for Quantum and Optical Science, Swinburne University of Technology, Melbourne, Victoria, 3122, Australia}
\author{Gao Xianlong}
\affiliation{Department of Physics, Zhejiang Normal University, Jinhua 321004, China}
\date{\today}

\begin{abstract}
We theoretically investigate one-dimensional three-component spin-orbit-coupled Fermi gases in the presence of Zeeman field. By solving the Bogoliubov-de Gennes equations, we obtain the phase diagram at a given chemical potential and order parameter. We show that, with increasing the intensity of the Zeeman field, in addition to undergoing a phase transition from Bardeen-Cooper-Schrieffer (BCS) superfluid to topological superfluid, similar to the two-component system, the three-component system may exhibit some other interesting topological phase transitions. For example, by appropriately adjusting the chemical potential $\mu$, the system can be in a non-trivial topological superfluid in the whole region of the Zeeman field $h$. It also may initially be a topological superfluid and then translates to a topologically trivial BCS superfluid with increasing the field $h$. Even more exotically, the system may exhibit a re-entrance behavior, being a topological superfluid at small and large fields but a topologically trivial BCS superfluid in between at a mediate Zeeman field. It can therefore have two regions with zero-energy Majorana fermions. As a consequence of these interesting topological phase transitions, the system of the three-component spin-orbit coupled Fermi gases in a certain parameter range is more optimizing for the experimental realization of the topological phase due to the smaller magnetic field needed. Thus, a promising candidate for the realization of the topological phase is proposed.
\end{abstract}


\maketitle

\section{Introduction}
Exploring new states of matter is the eternal theme in condensed matter physics. In recent years, a new state of matter named topological superfluid has
attracted intense attentions. Similar to topological insulators \cite{01,02}, they are fully gapped in the bulk and yet possess gapless exotic excitations on the edges called the Majorana fermions \cite{03,04,05,06,07}, being their own antiparticles and obeying non-Abelian statistics.  It is shown that they can be used for topological quantum computation since they are robust against local perturbations \cite{08,09}. As a result, the experimental realization of the topological superfluid and associated Majorana fermions has become a desirable task and feasible proposals from the side of the theorists are greatly needed.

In condensed matter physics, there are many proposals on realizing a topological superfluid including making use of the proximity effect between an s-wave superconductor and the surface of a topological insulator \cite{10,11,12}, a two-dimensional (2D) p-wave pairing superconductor \cite{13,14}, InAs nanowires, and banded carbon nanotubes in one-dimensional (1D) systems \cite{15,16,17,18}. In those proposals, spin-orbit (SO) coupling plays a key role. Most recently, synthetic spin-orbit coupling has been realized in cold-atom experiments \cite{19,20} based on a Raman technique pioneered by Splieman's group \cite{21,22}. After that, many schemes were proposed on the related topics, such as 2D atomic Fermi gases with a strong Rashba SO coupling \cite{23,24,25,26} and trapped 1D SO-coupled fermionic atoms \cite{27,28,29,30}.

Unlike the spin-orbit coupling in condensed matter system, where the spin of an electron only has two components: spin up $\mid\uparrow\rangle$ and spin down $\mid\downarrow\rangle$, the (pseudo) spins in cold-atom physics are actually hyperfine states of an atom such as ${}^{40}\textrm{K}$ and ${}^{6}\textrm{Li}$. Thus, one can prepare the superfluid by the s-wave Feshbach resonance with more than two hyperfine states~\cite{31,32,33}. Motivated by the above observation, in this paper, we discuss the superfluid with three (pseudo) spins and concentrate on the potential topological properties in the three-component Fermi gases. Within the mean-field approximation, we find that, in a certain parameter regime, the topological superfluid in the three-component Fermi gases can be achieved with a smaller magnetic field than the two-component system.

This paper is organized as follows. In Sec.~\uppercase\expandafter{\romannumeral2}, we discuss the model Hamiltonian. We solve the Bogoliubov-de-Gennes (BdG) equation and obtain the dispersion relation. The Berry phase is defined to describe the topology of the three-component superfluid. The Majorana zero modes and localized wave functions of the system are described. In Sec.~\uppercase\expandafter{\romannumeral3}, we present the $\mu-h$ phase diagram at a given order parameter, where $\mu$ is the chemical potential of the system and $h$ the magnetic field. In Sec.~\uppercase\expandafter{\romannumeral4}, we consider the spin-dependent situation with two different order parameters  $\Delta_{0,-1}=\Delta_{1,0}=\Delta$ and $\Delta_{1,-1}$, then we obtain the equation of phase transition. Finally, in Sec.~\uppercase\expandafter{\romannumeral5}, we present our conclusion.

\section{Model Hamiltonian}
We consider the Hamiltonian of 1D three-component SO coupled Fermi gases under Zeeman field,
\begin{eqnarray}
H& &=\int \Psi^{\dag}(x)[H_{0}(x)-\mu]\Psi(x) dx +\textsl{H}_{\rm I},\nonumber\\
\textsl{H}_{\rm I}&&=\textsl{g}_{1,0} \int \psi^{\dagger}_{1} (x)\psi^{\dagger}_{0}(x)\psi_{0} (x)\psi_{1}(x)dx \nonumber\\
& &+ \textsl{g}_{1,-1} \int \psi^{\dagger}_{1} (x)\psi^{\dagger}_{-1}(x)\psi_{-1} (x)\psi_{1}(x)dx \nonumber\\
& &+ \textsl{g}_{0,-1} \int \psi^{\dagger}_{0} (x)\psi^{\dagger}_{-1}(x)\psi_{-1} (x)\psi_{0}(x)dx,
\end{eqnarray}
where $\Psi^{\dagger}(x)=[\psi^{\dagger}_{1}(x),\psi^{\dagger}_{0}(x),\psi^{\dagger}_{-1}(x)]$
is the creation operator for the three hyperfine states. $\mu$ is the chemical potential. $H_I$ denotes the attractive interaction between the states $|1\rangle$ and $|0\rangle$, $|1\rangle$ and $|-1\rangle$, and $|0\rangle$ and $|-1\rangle$ with coupling constant $ \textsl{g}_{i,j}=-2\hbar^{2}/ma_{{\rm 1D}, ij}$ ($i,j=1,0,-1$). Here $ a_{{\rm 1D}, ij} $ is the 1D \textsl{s}-wave scattering length. A promising candidate of our proposed system is a three-component Fermi gas of ${}^{6}\textrm{Li}$ atoms near the broad Feshbach resonance 834 G trapped in a two-dimensional optical lattice \cite{31}. The single-particle Hamiltonian $ \textsl{H}_{0}(x)=-\frac{\hbar^{2}}{2m}\partial^{2}_{x}-\frac{2i\hbar^{2}k_{r}}{m}\partial_{x}\varsigma_{z}+4E_{r}\varsigma_{z}^{2}+h\varsigma_{x}+\delta_{1}\varsigma_{1}-\delta_{2}\varsigma_{2}$ can be experimentally realized \cite{19,20}. Here, $ h=\frac{\hbar\Omega}{2} $ denotes the strength of the Zeeman field with $\Omega$ is the Rabi frequency characterizing the intensity of the Raman laser, $ E_{r}=\frac{\hbar^{2}k_{r}^{2}}{2m} $  the recoil energy, and $ \hbar k_{r} $ the associated recoil momentum of the Raman laser. $\delta_{1}=\Delta_{1}+\Delta_{2}+\delta\omega$ and $\delta_{2}=\Delta_{1}-\Delta_{2}+\delta\omega$, where $\Delta_{1}$ and $\Delta_{2}$ are the linear and quadratic Zeeman energy, $\delta\omega$ denotes the frequency difference of the two Raman lasers. $ \varsigma_{x} $, $ \varsigma_{z}$, $ \varsigma_{1} $ and $ \varsigma_{2}$ are matrices defined as,
\begin{eqnarray}
 \varsigma_{x}&&=\left(
   \begin{array}{ccc}
     0 & 1 & 0 \\
     1 & 0 & 1 \\
     0 & 1 & 0 \\
   \end{array}
 \right),~~~~~~
 \varsigma_{z}=\left(
                 \begin{array}{ccc}
                   1 & 0 & 0 \\
                   0 & 0 & 0 \\
                   0 & 0 & -1 \\
                 \end{array}
               \right),\nonumber \\
\varsigma_{1}&&=\left(
   \begin{array}{ccc}
     1 & 0 & 0 \\
     0 & 0 & 0 \\
     0 & 0 & 0 \\
   \end{array}
 \right),~~~~~~
 \varsigma_{2}=\left(
                 \begin{array}{ccc}
                   0 & 0 & 0 \\
                   0 & 0 & 0 \\
                   0 & 0 & 1 \\
                 \end{array}
               \right)~.
\end{eqnarray}
By adjusting the values of $\delta_{1}$ and $\delta_{2}$ appropriately, one can reach the three-minima regime experimentally as discussed in \cite{33}. Since the effects of $\delta_{1}$ and $\delta_{2}$ are negligible, for the sake of simplicity, we set $\delta_{1}=\delta_{2}=0$.

Within the mean-field approximation, the interaction term becomes,
\begin{eqnarray}
\textsl{H}_{\rm I}\thickapprox \int \big(&&\Delta_{1,0}(x)\psi^{\dagger}_{0}(x) \psi^{\dagger}_{1}(x)+ \Delta_{1,-1}(x)\psi^{\dagger}_{1}(x) \psi^{\dagger}_{-1}(x)
\nonumber\\
+&&\Delta_{0,-1}(x)\psi^{\dagger}_{-1}(x) \psi^{\dagger}_{0}(x)+ H.c.\big)~dx.
\end{eqnarray}
Here
\begin{eqnarray}
 \Delta_{1,0}(x)&&= \textsl{g}_{1,0}\langle\psi_{0}(x) \psi_{1}(x)\rangle, \nonumber \\
 \Delta_{1,-1}(x)&&= \textsl{g}_{1,-1}\langle\psi_{1}(x) \psi_{-1}(x)\rangle,\nonumber \\
 \Delta_{0,-1}(x)&&= \textsl{g}_{0,-1}\langle\psi_{-1}(x) \psi_{0}(x)\rangle,
 \end{eqnarray}
 are the order parameters. Without loss of generality and aiming at a clear phase diagram, we mainly consider the homogeneous situation and assumed it to be real, i.e., $\Delta_{i,j}(x)=\Delta,(i,j=1,0,-1)$, which may be experimentally achieved by exploiting the proximity effect~\cite{34}.

\subsection{The BdG equation and dispersion relation}
To study the basic properties of the 1D three-component superfluid with spin-orbit coupling, we calculate the elementary excitations by solving the BdG equation\cite{23,25,26}, which reads,
\begin{equation}
\textsl{H}_{\rm BdG}\Psi_{\eta}(x)=\textsl{E}_{\eta}\Psi_{\eta}(x),
\end{equation}
where $\Psi_{\eta}(x)=[u_{1,\eta}(x),u_{0,\eta}(x),u_{-1,\eta}(x),
v_{1,\eta}(x),v_{0,\eta}(x),\\v_{-1,\eta}(x)]^{\textsl{T}}$ is spinor in Nambu representation and the Hamiltonian $ \textsl{H}_{\rm BdG}$ is,
\begin{widetext}
\begin{equation}
\textsl{H}_{\rm BdG}=\left(
          \begin{array}{cccccc}
            \textsl{H}_{\rm s}^{\textsl{0}}-\frac{2i\hbar^{2}k_{r}}{m}\partial_{x}+4E_{r} & h & 0 & 0 & -\Delta & \Delta \\
            h & \textsl{H}_{\rm s}^{\textsl{0}} & h & \Delta & 0 & -\Delta \\
            0 & h & \textsl{H}_{\rm s}^{\textsl{0}}+\frac{2i\hbar^{2}k_{r}}{m}\partial_{x}+4E_{r}& -\Delta& \Delta & 0 \\
            0 & \Delta & -\Delta & -\textsl{H}_{\rm s}^{\textsl{0}}-\frac{2i\hbar^{2}k_{r}}{m}\partial_{x}-4E_{r} & -h & 0 \\
            -\Delta & 0 & \Delta & -h & -\textsl{H}_{\rm s}^{\textsl{0}} & -h \\
             \Delta& -\Delta & 0 & 0 & -h & -\textsl{H}_{\rm s}^{\textsl{0}}+\frac{2i\hbar^{2}k_{r}}{m}\partial_{x}-4E_{r} \\
          \end{array}
        \right),
\end{equation}
\end{widetext}
where $ \textsl{H}_{\rm s}^{\textsl{0}}=-\frac{\hbar^{2}}{2m}\partial^{2}_{x}-\mu$. We immediately note that, by introducing the Bogoliubov transformation \cite{35},
\begin{equation}
\psi_{i}(x)=\sum_{\eta,j=-1,0,1}~[u_{i,\eta}(x)\Gamma_{j,\eta}+v_{i,\eta}^{*}(x)\Gamma_{j,\eta}^{\dag}],
\end{equation}
similar to the two-component case, the three-component BdG Hamiltonian exhibits the particle-hole redundancy as well \cite{26}.
Therefore, the associated Bogoliubov quasiparticle operators satisfy $ \Gamma_{\rm E}=\Gamma^{\dagger}_{- \rm E}$.

\begin{figure}
  \centering
  \includegraphics[width=0.5
  \textwidth]{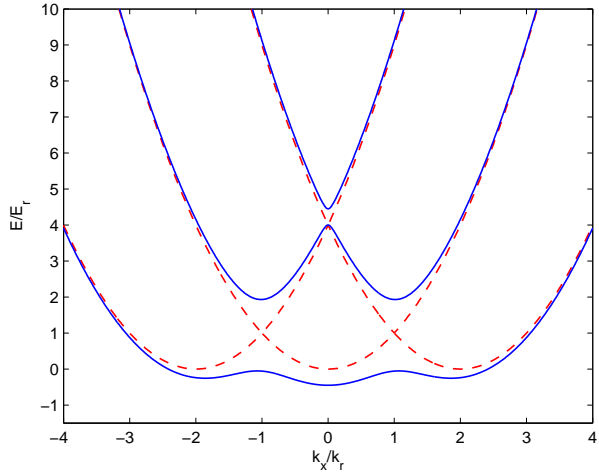}\\
  \caption{(Color online) The single-particle band structure of a three-component Fermi gas. The (red) dashed lines are three bands in the absence of the Zeeman field $ h/E_{r}=0$, while the (blue) solid lines refer to $ h/E_{r}=1$.}
  \label{fig:threeband}
\end{figure}

\begin{figure}
  \centering
  \includegraphics[width=0.5\textwidth]{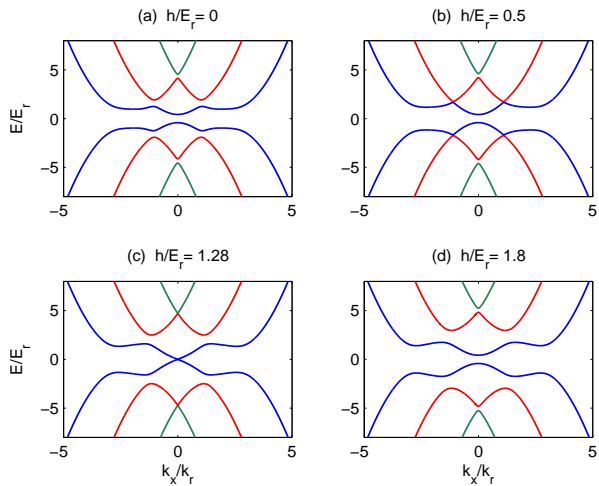}\\
  \caption{(Color online) Spectrum in momentum space for the homogeneous system. The parameters are $\mu=0$ and $\Delta=E_{r}$. As the Zeeman field increases, we see the gap closes at $ k_{x}=0$ when $h_c=1.28E_{r}$ and then reopens for $h>h_c$.}
\end{figure}

The physics of the single-particle Hamiltonian is simple in momentum space : $ \textsl{H}_{0}(k)=\frac{\hbar^{2}k_{x}^2}{2m}-\frac{2\hbar^{2}k_{r}k_{x}}{m}\varsigma_{z}+4E_{r}\varsigma_{z}^{2}+h\varsigma_{x}$. The three-band structure of a three-component Fermi gas is illustrated in Fig. \ref{fig:threeband}.
Since we consider in this paper the homogeneous system without external potential, one can write the above BdG equation in momentum space as $ H_{k}W(k)=E(k)W(k)$ easily, by replacing $-i\partial_{x}\rightarrow k$ in Eq. (5). After diagonalizing the Hamiltonian $ H_{k}$ we numerically get the dispersion relation. In Fig.~2, we show the spectrum at the fixed chemical potential $\mu=0$ and the order parameter $\Delta=E_{r}$. As the Zeeman field $h$ increases, the system evolves from the a conventional BCS superfluid $(h<h_{c})$ to a topological superfluid $(h>h_{c})$. Here the critical Zeeman field $h_{c}=1.28E_{r}$, which is different from the two-component case as given by $ h_{c}= \sqrt{\mu^{2}+\Delta^{2}}=1 E_r$.

\subsection{Majorana zero modes and wave functions}
The emergence of Majorana fermions in a spin-orbit-coupled Fermi gas \cite{23,24,25,26,28,29} is a very attractive feature because of their non-Abelian statistics. Since a Majorana fermion is its own antiparticle and there exists an inherent particle-hole redundancy in the BdG equation ($ \Gamma_{\rm E}=\Gamma^{\dagger}_{-\rm E}$), we know from the 1D two-component Fermi gases with SO coupling, zero-energy Majorana modes ($\Gamma_{0}^{\dagger}=\Gamma_{0}$) will emerge when the superfluid becomes topological non-trivial. Furthermore, the wave functions of these Majorana fermions are localized at the two edges of one-dimensional system. We note that, for the three-component case, the similar topological transition will be observed when adjusting the intensity of Zeeman field $h$ in a certain chemical potential regime. However, three new interesting phase regions appear in the parameter space which will be addressed in the following phase diagram.

To illustrate the topological properties of Fermi gases, one needs to calculate the Berry phase~\cite{29}, which is defined as,
\begin{eqnarray}
\gamma=&&~i\oint_{c}\textsl{W}^{*}\cdot\partial_{k}\textsl{W}\\
=&&~i\int^{\infty}_{-\infty}[\textsl{W}^{~1}(k)]^{*}\cdot\partial_{k}\textsl{W}^{~1}(k) \textsl{dk}\nonumber\\
&&+~i\int^{\infty}_{-\infty}[\textsl{W}^{~0}(k)]^{*}\cdot\partial_{k}\textsl{W}^{~0}(k) \textsl{dk}\nonumber\\
&&+~i\int^{\infty}_{-\infty}[\textsl{W}^{~-1}(k)]^{*}\cdot\partial_{k}\textsl{W}^{~-1}(k) \textsl{dk},
\end{eqnarray}
in the present three-component Fermi gases. Here $\textsl{W}^{~1}(k), \textsl{W}^{~0}(k)$ and, $
\textsl{W}^{~-1}(k)$ are three eigenvectors of $H_{k}$ corresponding to the three lower (hole) bands. For the sake of smoothness, instead, we calculate
\begin{eqnarray}
e^{i\gamma}=&&\lim_{\delta k\rightarrow 0}\bigg(\prod_{k=-\infty}^{\infty}[\textsl{W}^{~1}(k)]^{*}\cdot\textsl{W}^{~1}(k-\delta k)\nonumber\\
&& \times \prod_{k=-\infty}^{\infty}[\textsl{W}^{~0}(k)]^{*}\cdot\textsl{W}^{~0}(k-\delta k)\nonumber\\
&& \times \prod_{k=-\infty}^{\infty}[\textsl{W}^{~-1}(k)]^{*}\cdot\textsl{W}^{~-1}(k-\delta k)\bigg)\nonumber\\
&& \times \bigg([\textsl{W}^{~1}(-\infty)]^{*}\cdot\textsl{W}^{~-1}(\infty)\nonumber\\
&& \times[\textsl{W}^{~-1}(-\infty)]^{*}\cdot\textsl{W}^{~1}(\infty)\nonumber\\
&& \times[\textsl{W}^{~0}(-\infty)]^{*}\cdot\textsl{W}^{~0}(\infty)\bigg).
\end{eqnarray}

As a result of real wave-functions for the Hamiltonian, it is easy to see that $e^{i\gamma}$ can be either $+1$ or $-1$\cite{29}. The above definition for the Berry phase is similar to that for a 1D topological insulator in expression~\cite{01,02}. However, there is an important difference. In 1D topological insulators, the integration over the Berry connection is closed at the first Brillouin zone, where the two boundaries are identified as the same point (i.e., periodic condition). In our continuous system, the two boundaries are instead given by $k=\pm \infty$. Thus, the Berry connections at the two boundaries may take the same value only up to a phase factor. As a result of this ambiguity in phase factor, as we shall see below, somehow counterintuitively the topological trivial and non-trivial phases are characterized by $e^{i\gamma}=-1$ and $e^{i\gamma}=+1$, respectively. This was pointed out earlier by Wei and Mueller in Ref. \cite{29} for a two-component Fermi system.

\begin{figure}
  \centering
  \includegraphics[width=0.5\textwidth]{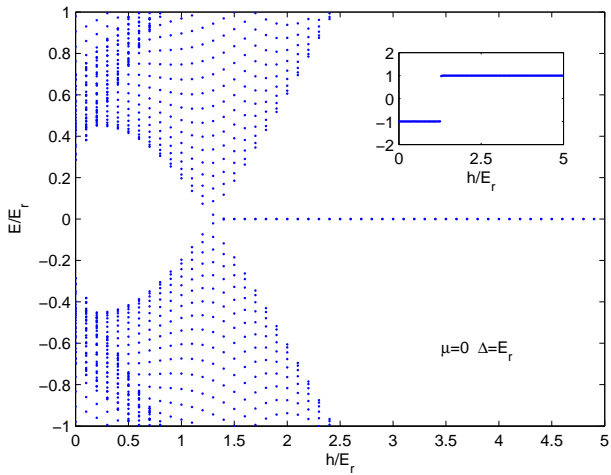}\\
  \caption{(Color online)  Profile of energy spectrum for the system of open boundary conditions. Zero mode emerges at Zeeman field $h=1.4E_{r}$. The inset shows Berry phase $e^{i\gamma}$ varies from $-1$ to $1$ as  the system evolves from the conventional superfluid to the topological superfluid.}
\end{figure}

\begin{figure}
  \centering
  \includegraphics[width=0.5\textwidth]{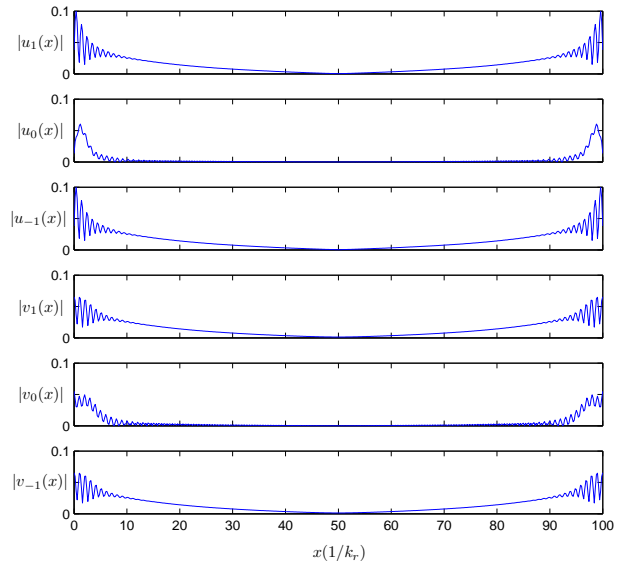}\\
  \caption{(Color online)  Wave functions for different components with the system of length $x=100/k_{r}$. By plotting the square root of the wave function $\Psi_{\eta}(x)$, we see that all these wave functions are localized at two edges. Here $|u_{i}(x)|$ and $|v_{i}(x)|$ refer to the particle and hole part, respectively. From Eq~(7), we know that, the Bogoliubov transformation is a combination of quasiparticle operators $\Gamma_{-1,\eta},\Gamma_{0,\eta}$ and $\Gamma_{1,\eta}$. As a result, the Majorana modes are manifested in $|u_{i}(x)|$ and $|v_{i}(x)|,(i=-1,0,1)$.}
\end{figure}
 
In Fig.~3, we show the numerical result of energy spectrum with open boundary condition under the fixed chemical potential $\mu=0$ and order parameter $\Delta=E_{r}$. The zero mode emerges when Zeeman field $h$ exceeds $h_{c}=1.4E_{r}$. The inset on the top shows the Berry phase evolves from $e^{i\gamma}=-1$ to $e^{i\gamma}=1$ (where the system exhibits the zero-energy modes and localized wave functions as indicated in Figs.~3 and 4, respectively) as increasing $h $, which indicates the system undergoes a topological transition from the topological trivial superfluid phase to the topological nontrivial superfluid phase. The wave functions for different components are depicted in Fig.~4 when $h=3E_{r}$, which shows the localized characteristic of the Majorana fermionic states.

\begin{figure}
  \centering
  \includegraphics[width=0.5\textwidth]{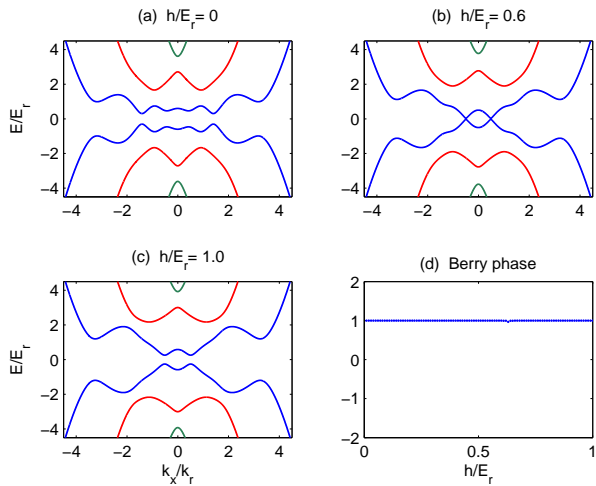}\\
  \caption{(Color online) Band structure of the homogeneous gas under fixed parameters $\mu=1.5E_{r}$ and $\Delta=E_{r}$. (a)-(c) shows the spectrum at $h=0$, $h=0.6E_{r}$, and $h=E_{r}$, respectively; (d) illustrates that the Berry phase keeps the same even when the gap closes at the nonzero $k_x/k_r$. }
\end{figure}

\section{Phase diagram}

In this section we report the phase diagram of the three-component system. Firstly, it should be pointed out that, unlike the situation shown in Fig.~2, where the system undergoes a topological transition when the middle two bands (blue lines) close at $k_{x}=0$, for certain parameter regime, such as for $\mu=1.5E_{r}$ and $\Delta=E_{r}$, the gap closes at $k_{x}\neq 0$ (see Fig. 5(b)). We calculate the Berry phase, it turns out that the system does not change its topological property at these gapless points, as shown in Fig. 5(d). Therefore, we believe that the Berry phase changes only when gapless spectrum emerges at $k_{x}=0$. By analytically calculating the energy spectrum at $k_x=0$,  we find that the phase of the system is governed by a simple analytical equation in terms of the chemical potential $\mu$, the order parameter $\Delta$ and the critical Zeeman field $h_{c}$,
\begin{equation}
Ah_{c}^{2}+Bh_{c}+C=0,
\end{equation}
with $A=2\mu-8, B=4\Delta^{2}, C=8(\Delta^{2}+\mu^{2})-3\Delta^{2}\mu-16\mu-\mu^{3}$.
The solutions of the above equation are,
\begin{eqnarray}
&&h_{c}=1 ~~~~~~~~~~~~~~~~~~~~~~~~~~~~~~(\mu=4)~,\nonumber\\
&&h_{c}=\frac{-B\pm\sqrt{B^{2}-4AC}}{2A}~~~~~~(\mu\neq4)~,
\end{eqnarray}
with
\begin{equation}
B^{2}-4AC=8[\Delta^{2}+(\mu-4)^{2}][2\Delta^{2}-4+(\mu-2)^{2}].
\end {equation}
This equation is obtained by the following steps : firstly, write down the BdG equation in momentum space as $ H_{k}W(k)=E(k)W(k)$, to get the eigenvalues one should calculate the determinant

\begin{widetext}
\begin{equation}
\left|
  \begin{array}{cccccc}
    \textsl{H}_{\rm k}^{\textsl{0}}+\frac{2\hbar^{2}k_{r}k_{x}}{m}+4E_{r} & h & 0 & 0 & -\Delta & \Delta \\
    h & \textsl{H}_{\rm k}^{\textsl{0}} & h & \Delta & 0 & -\Delta \\
    0 & h & \textsl{H}_{\rm k}^{\textsl{0}}-\frac{2\hbar^{2}k_{r}k_{x}}{m}+4E_{r} & -\Delta & \Delta & 0 \\
    0 & \Delta & -\Delta & -\textsl{H}_{\rm k}^{\textsl{0}}+\frac{2\hbar^{2}k_{r}k_{x}}{m}-4E_{r} & -h & 0 \\
    -\Delta & 0 & \Delta & -h & -\textsl{H}_{\rm k}^{\textsl{0}} & -h \\
    \Delta & -\Delta & 0 & 0 & -h & -\textsl{H}_{\rm k}^{\textsl{0}}-\frac{2\hbar^{2}k_{r}k_{x}}{m}-4E_{r}\\
  \end{array}
\right|=E,
\end{equation}
\end{widetext}
here $ \textsl{H}_{\rm k}^{\textsl{0}}=\frac{\hbar^{2}k_{x}^{2}}{2m}-\mu$; secondly, let $ k_{x}$ and $E$ to be zero, since we assume the system evolves a topological phase transition only when the middle two bands (blue lines) are closed and $k_{x}=0$ simultaneously, as discussed before; finally, expand the determinant then we get the above equation.

\begin{figure}
  \centering
  \includegraphics[width=0.5\textwidth]{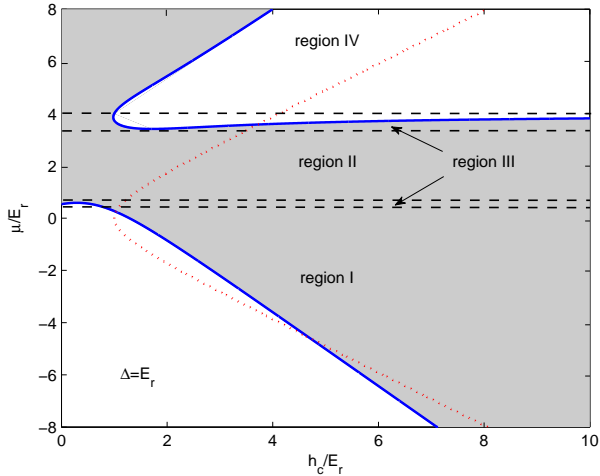}\\
  \caption{(Color online)  Phase diagram at $\Delta=E_{r}$. Comparing to the two-component superfluid (red dotted line), the three-component case (blue lines) is more complicated. Four regions are divided by black dashed lines at different chemical potentials $\mu$. The grey area shows the topologically non-trivial phase. The details of the region I to IV are explained in the main text.}
\end{figure}

\begin{figure}
  \centering
  \includegraphics[width=0.5\textwidth]{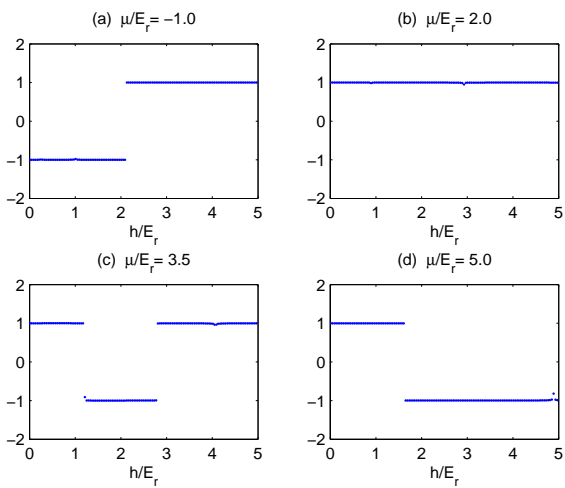}\\
  \caption{(Color online) Profile of the Berry phase $e^{i\gamma}$ at the fixed order parameter $\Delta=E_{r}$ with different $\mu$. (a)-(d) is corresponding to the region I-IV shown in Fig.~6, respectively.}
\end{figure}

\begin{figure}
  \centering
  \includegraphics[width=0.5\textwidth]{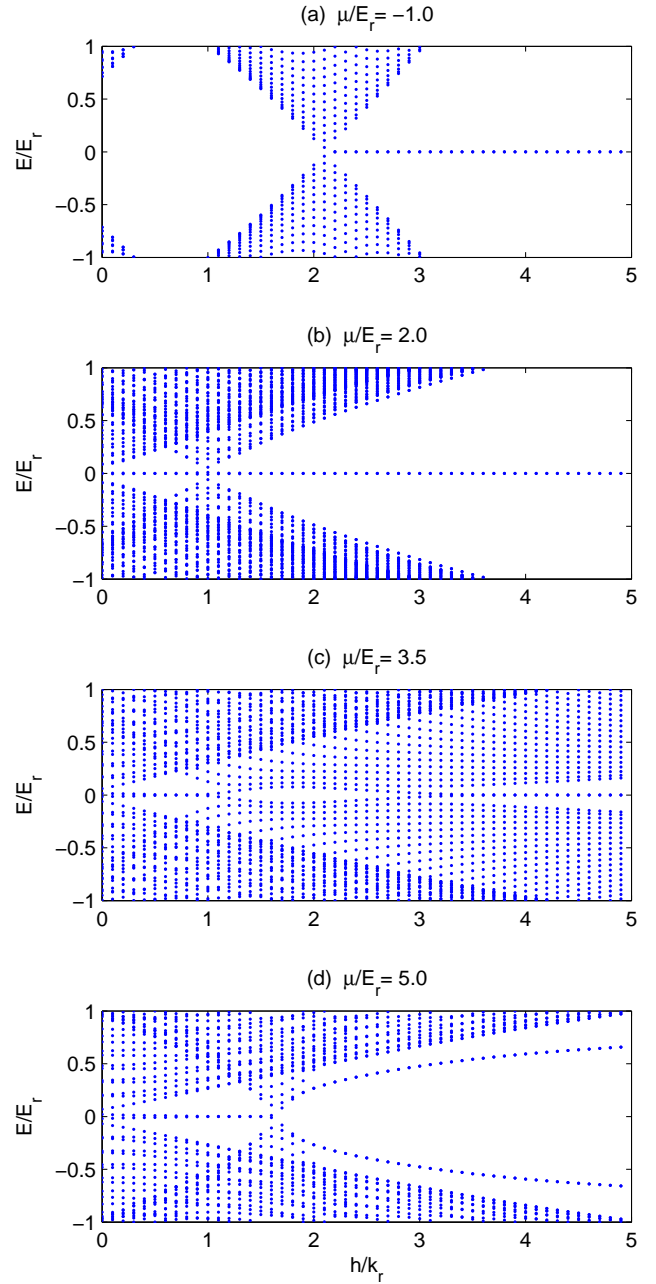}\\
  \caption{(Color online) Profile of the energy spectrum at four different phase regions. The parameters are the same as those in Fig.~7. (a) $\mu=-E_{r}$. The system undergoes a 'conventional' topological phase transition when exceeds the critical Zeeman field $h_{c}=2.2E_{r}$; (b) $\mu=2E_{r}$. The system is topological non-trivial at any values of $h$; (c) $\mu=3.5E_{r}$. It has two critical Zeeman fields with $h_{c1}=E_{r}$ and $h_{c2}=3.3E_{r}$. The system is topological trivial when $h_{c1}<h<h_{c2}$; (d) $\mu=5E_{r}$. The zero mode disappears when $h>h_{c}=1.6E_{r}$. }
\end{figure}

In Fig.~6 we plot the phase diagram of three-component superfluid (blue solid lines) at the given order parameter $\Delta=E_{r}$. For comparison, the phase diagram of the two-component superfluid (red dotted line) is plotted as well. Since $ h=\frac{\hbar\Omega}{2} $ refers to the intensity of the Raman laser, it should be real and positive. One can distinguish four different phase regions in the $\mu-h$ phase diagram, say region I-IV, according to the different behaviors of the Berry phase. In Figs.~7 and 8, we plot the Berry phase and the zero energy modes corresponding to the different regions discussed in the phase diagram as the function of the Zeeman field. In the following part, we discuss in details the four different phase regions.

Region I ($\mu\leq\mu_{c1}=0.53E_{r}$). The phase transition behavior of the system in this region is much similar to the conventional two-component superfluid case, i.e., there exists a critical Zeeman field $ h_{c}$ and the system undergoes a topological phase transition, from a topological trivial ($ h<h_{c}$) to a topological nontrivial ($ h>h_{c}$) phase. For instance, when $\mu=-E_{r}$, the critical Zeeman field is $h_{c}=2.2E_{r}$ (The analytical result from Eq. (10) gives $h_{c}=2.11E_{r}$. The deviation is due to the finite system we studied. If one increases the length of system, this problem will be mitigated effectively). The corresponding profiles of the berry phase and the energy spectrum in region I are illustrated in Figs.~7(a) and 8(a). The phase boundaries for the two- and three-component system in this region intersect at  $\mu_{1}=-4.59E_{r}$ and $\mu_{2}=0.24E_{r}$. We can conclude that at $ \mu<\mu_{1}$ or $\mu>\mu_{2}$, the critical magnetic field for the topological phase transition of the three-component system is smaller than that of the two-component one. That is, it will become easier to realize this interesting topological phase with smaller magnetic field, which is the necessary condition for keeping the superfluid gap finite.

Region II ($ \mu\in[\mu_{c2},\mu_{c3}]$, with $\mu_{c2}=0.59E_{r}$ and $\mu_{c3}=3.41E_{r}$). In this region,  $B^{2}-4AC<0$. There is no solution for $h_c$. As a result, the system has no topological transition at any values of  $h$. To judge the topological characteristic of this region, one should calculate the Berry phase. We find that the system is totally topological non-trivial even when $h=0$. In Figs.~7(b) and 8(b), we plot the berry phase and the energy spectrum to illustrate the topological characteristic of system when $\mu=2E_{r}$. It should be emphasized that the range of region II is highly dependent on the value of order parameter $\Delta$. One can easily find that at $\Delta=0$, the range of region II is $\mu\in(0,4E_{r})$. However, the increasing of the intensity of the order parameter $\Delta$ rapidly shrinks this region and it disappears when $\Delta$ exceeds $\Delta_{c}=\sqrt{2}E_{r}$.

Region III ($ \mu\in[\mu_{c1},\mu_{c2}]\cup[\mu_{c3},\mu_{c4}]$ with $\mu_{c4}=4E_{r}$). In this region, there exists two critical Zeeman fields $h_{c1}$ and $h_{c2}$. Unlike the conventional phase transition as shown in Region I, the system is topological non-trivial at first, as increasing the Zeeman field $h$ it becomes topological trivial (for $h\in[h_{c1}$, $h_{c2}]$) and finally it becomes non-trivial again (for $h>h_{c2}$). When $\mu=3.5E_{r}$ the corresponding berry phase and the energy spectrum behavior in real space are plotted in Figs.~7(c) and 8(c) respectively. From  Fig.~8(c),  we see that as $h$ varying it has two rather than one zero-mode regions.

Region IV ($ \mu\geq\mu_{c4}=4E_{r}$). Comparing with the region I, it presents a reversed topological behavior, i.e., the system is topological trivial at the small magnetic field $h$ and it becomes topological nontrivial when $h>h_{c}$. The topological transition behavior in this region for $\mu=5E_{r}$ is illustrated in the corresponding Berry phase in Fig.~7(d) and energy spectrum in Fig.~8(d), where the zero modes disappear at $h_c=1.6E_{r}$.

\begin{figure}
  \centering
  \includegraphics[width=0.5\textwidth]{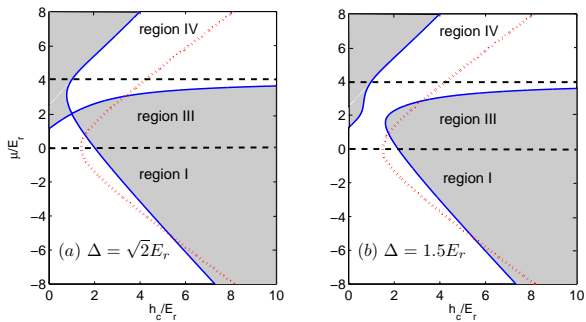}\\
  \caption{(Color online) Phase diagram at (a) $\Delta=\sqrt{2}E_{r}$ and (b) $\Delta=1.5E_{r}$, respectively. When $\Delta>\sqrt{2}E_{r}$, it has only three regions in this situation. The (blue) lines are for the three-component superfluid while the (red) dotted line for the two-component one. The topologically non-trivial phase is highlighted by the grey area.}
\end{figure}

 At last, we discuss the phase transitions for larger $\Delta$ ($>E_r$). We plot the phase diagram at $\Delta=\sqrt{2}E_{r}$ and $\Delta=1.5E_{r}$ in Fig.~9. As we discussed before, when $\Delta >\sqrt{2}E_{r}$, the inequality $B^{2}-4AC>0$ holds for any $\mu$, thus the region II in Fig.~6 disappears and there are only three different phase regions in this situation.

It should be pointed out that, for three-component superfluid, there exists topological non-trivial phases even in the absence of spin-orbit coupling $(h=0)$ as long as the system is in regions II, III or IV, which is quite different from the two-component case. The reason is that, for the three-component system, the three order parameters $\Delta_{0,-1}$, $\Delta_{1,0}$ and $\Delta_{1,-1}$ do not satisfy gauge invariant simultaneously. To prove it, we introduce the following gauge transformation,
\begin{equation}
\phi_{-1}=\psi_{-1} e^{2ik_{r}x};~~~~~~\phi_{0}=\psi_{0};~~~~~~\phi_{1}=\psi_{1} e^{-2ik_{r}x}.
\end{equation}
One can easily show that, the Hamiltonian of system and $\Delta_{-1,1}$ are invariant after such gauge transformation while the order parameters $\Delta_{0,-1}$, $\Delta_{1,0}$ are gauge-dependent, that is,
\begin{align}
&\Delta_{0,-1}^{\phi}= \textsl{g}_{\rm 1D}\langle\phi_{0} \phi_{-1}\rangle=\textsl{g}_{\rm 1D}\langle\psi_{0} \psi_{-1}\rangle e^{2ik_{r}x}=\Delta_{0,-1}^{\psi}e^{2ik_{r}x},\nonumber\\
&\Delta_{1,0}^{\phi}= \textsl{g}_{\rm 1D}\langle\phi_{1} \phi_{0}\rangle=\textsl{g}_{\rm 1D}\langle\psi_{1} \psi_{0}\rangle e^{-2ik_{r}x}=\Delta_{1,0}^{\psi}e^{-2ik_{r}x},\nonumber\\
&\Delta_{1,-1}^{\phi}= \textsl{g}_{\rm 1D}\langle\phi_{1} \phi_{-1}\rangle=\textsl{g}_{\rm 1D}\langle\psi_{1} \psi_{-1}\rangle =\Delta_{1,-1}^{\psi}.
\end{align}
Now it becomes clear that, due to the inherent gauge-symmetry breaking, the system has the potential to exhibit topological non-trivial phases even without spin-orbit coupling.

\section{Spin-dependent situation}
In this section we extend our discussion to the spin-dependent situation. As we discussed in Sec.~\uppercase\expandafter{\romannumeral3}, since $\Delta_{0,-1}$ and $\Delta_{1,0}$ are gauge-dependent while $\Delta_{-1,1}$ is gauge-independent, we set $\Delta_{0,-1}=\Delta_{1,0}=\Delta$. Then the BdG equation in momentum space should be modified as
\begin{widetext}
\begin{equation}
\left|
  \begin{array}{cccccc}
    \textsl{H}_{\rm k}^{\textsl{0}}+\frac{2\hbar^{2}k_{r}k_{x}}{m}+4E_{r} & h & 0 & 0 & -\Delta & \Delta_{1,-1} \\
    h & \textsl{H}_{\rm k}^{\textsl{0}} & h & \Delta & 0 & -\Delta \\
    0 & h & \textsl{H}_{\rm k}^{\textsl{0}}-\frac{2\hbar^{2}k_{r}k_{x}}{m}+4E_{r} & -\Delta_{1,-1} & \Delta & 0 \\
    0 & \Delta & -\Delta_{1,-1} & -\textsl{H}_{\rm k}^{\textsl{0}}+\frac{2\hbar^{2}k_{r}k_{x}}{m}-4E_{r} & -h & 0 \\
    -\Delta & 0 & \Delta & -h & -\textsl{H}_{\rm k}^{\textsl{0}} & -h \\
    \Delta_{1,-1} & -\Delta & 0 & 0 & -h & -\textsl{H}_{\rm k}^{\textsl{0}}-\frac{2\hbar^{2}k_{r}k_{x}}{m}-4E_{r}\\
  \end{array}
\right|=E~.
\end{equation}
\end{widetext}

By performing the similar procedures that we introduced in Sec.~\uppercase\expandafter{\romannumeral3}, we immediately obtain the equation of phase transitions,
\begin{equation}
Ah_{c}^{2}+Bh_{c}+C=0,
\end{equation}
here
\begin{eqnarray}
A=&&2\mu-8\nonumber\\
B=&&4\Delta\Delta_{1,-1}\nonumber\\
C=&&-\mu^{3}+8\Delta^{2}-16\mu+8\mu^{2}-2\Delta^{2}\mu-\Delta_{1,-1}^{2}\mu,
\end{eqnarray}
and
\begin{equation}
B^{2}-4AC=8[\Delta_{1,-1}^{2}+\mu^{2}-8\mu+16][\mu^{2}-4\mu+2\Delta^{2}].
\end{equation}
The topological phase transition behavior of the spin-dependent system resembles the spin-independent one since it obeys the similar analytical equation but with the different values of coefficients $B$ and $C$.

\section{Conclusion}
In this paper we have theoretically investigated the  one-dimensional three-component spin-orbit-coupled Fermi gases in the presence of the Zeeman field. We first show the dispersion relation of the quasiparticle by solving the BdG equations and discuss the Majorana zero energy modes and the localized wave functions. The Berry phase $e^{i\gamma}$ is calculated to distinguish the topological properties, with $e^{i\gamma}=-1$ corresponding to a BCS superfluid while $e^{i\gamma}=1$ refers to a topological superfluid. Finally we proposed an analytical equation of topological phase transition in terms of the critical Zeeman field $h_{c}$ at the given chemical potential $\mu$ and order parameter $\Delta$. According to this equation, we found that the phase diagram in a three-component system is much richer than the two-component one.

By analyzing the phase diagram carefully, we found there exists four different phase regions. One of them is similar to the two-component superfluid with the topological phase transition from the trivial superfluid to nontrivial topological superfluid as increasing the magnetic field. However there exists three other completely different topological regions, i.e., a region where the system is always in a nontrivial topological superfluid, a region with two Majorana zero energy regions, and a region where the topological phase transition behaves reversely as the two-component system from the nontrivial topological superfluid to trivial superfluid as increasing the magnetic field.

By comparing to the two-component superfluid, we found the system we discussed is more optimizing for experimental realization in a certain parameter space due to the smaller magnetic field needed. We therefore expect that the three-component spin-orbit-coupled Fermi gases is a promising candidate for realizing topological superfluid.

Acknowledgements.
We acknowledge the useful discussions with Yi Wei. This work was supported by the NSF of China (Grant Nos. 11374266 and 11174253), the Zhejiang Provincial Natural Science Foundation (Grant No. R6110175), the Program for New Century Excellent Talents in University, and the ARC Discovery Projects (Grant Nos. FT130100815 and DP140103231).

\end{document}